\documentstyle[aps]{revtex}
\def\be{\begin{equation}}
\def\ee{\end{equation}}
\def\bi{\bibitem}
\begin{document}
\title{QUANTUM COSMOLOGY WITH $R + R^2$ GRAVITY} 
\draft
\author{{A.K.Sanyal}
\thanks{e-mail:aks@juphys.ernet.in}\\
Department of Physics, Jangipur College, Murshidabad, 742213\\
and\\ 
Relativity and Cosmology Research Centre, Department of Physics, Jadavpur 
University,Kolkata 700032\\ 
{B.Modak}
\thanks{e-mail:bmodak@klyuniv.ernet.in}\\
Department of Physics, Kalyani University, 741235, India\\}
\date{\today}
\maketitle
\begin{abstract}

Canonical quantization of an action containing curvature squared term 
requires introduction of an auxiliary variable. Boulware etal \cite{b:a} 
prescribed a technique to choose such a variable, by taking derivative of 
the action with respect to the highest derivative of the field variable, 
present in the action. It has been shown that \cite{a:p} this technique 
can even be applied in situations where introduction of auxiliary 
variables are not at all required, leading to wrong Wheeler-deWitt 
equation. It has also been pointed out that \cite{a:p} Boulware 
etal's \cite{b:a} prescription should be taken up only after removing all 
possible total derivative terms from the action. Once this is done only a 
unique description of quantum dynamics would emerge. For curvature 
squared term this technique yields, for the first time, a quantum 
mechanical probability interpretation of quantum cosmology, and an 
effective potential whose extremization leads to Einstein's equation. We 
conclude that Einstein-Hilbert action should essentially be modified by 
atleast a curvature squared term to get a quantum mechanical formulation of 
quantum cosmology and hence extend our previous work \cite{a:p} for such 
an action along with a scalar field.
\end{abstract}


\section{\bf{Introduction}}
Quantum cosmology is most elusive as the role of time \cite{k:q} is not 
unique and one is being unable to define the Hilbert space. This is 
due to the fact that `time' is not an external parameter in  
general theory of relativity, rather it is intrinsically contained in 
the theory, unlike its role in quantum mechanics or quantum field 
theory in flat space time. In curved space time, different slices 
correspond to different choices of time leading to inequivalent quantum 
theories. Likewise, the canonical quantization of gravity is devoid of 
an unique time variable and hence the definition of probability of 
emergence of a particular Universe out of an ensemble is ambiguous. 
The canonical quantization of Einstein-Hilbert action together with 
some matter fields yields the Wheeler-deWitt equation which does not 
contain time a priori, although it emerges intrinsically through the 
scale factor of the Universe. However, if the canonical variable is so 
chosen that one of the true degrees of freedom is disentangled 
from the kinetic part of the canonical variables, then this kinetic 
part in the corresponding quantum theory yields a quantum mechanical 
flavour of time. This is possible only if the Einstein-Hilbert action 
is replaced by curvature squared action or modified by the introduction 
of curvature squared term, in the Robertson-Walker minisuperspace model. 
In a recent publication \cite{a:p} we have presented such a 
choice of the canonical variable in a curvature squared action whose 
quantization yields a Schr$\ddot{o}$dinger like equation. Further, as a 
consequence, we could identify the nature of the space and time like 
variables in the Robertson-Walker minisuperspace model from the 
equation of continuity and hence could establish the idea of the 
probability and the current densities in the context of quantum 
cosmology.      
\par
The relevance of the fourth order gravity theory in cosmology was first 
revealed by Starobinsky \cite{s:p}, though it appeared earlier in the 
context of quantum field theory in curved space time. Starobinsky 
\cite{s:p} 
presented a solution of the inflationary scenerio without invoking phase 
transition in the very early Universe, from a field equation containing 
only the geometric terms. However, the field equations could not be 
obtained from the action principle, as the terms in the field equations 
are generated from the perturbative quantum field theory. Later, 
Starobinsky and Schmidt \cite{s:c} have shown that the inflationary phase 
is an attractor in the neighbourhood of the fourth order gravity theory.
Such wonderful relevance of higher order gravity theory in the context of 
cosmology inspired some authors to give interpretation of the quantum 
cosmological wavefunction with $R^2$ \cite{p:c} and even $R^3$ \cite {l:c} 
terms in the Einstein-Hilbert action. 
\par 
Further the functional integral for the wavefunction of the Universe 
proposed by Hartle-Hawking \cite{hh:p} runs into serious problem, since 
the wavefunction diverges badly. There are some prescriptions \cite{jj:p} to 
avoid 
such divergences, though a completely satisfactory result has not yet 
been obtained. However, to get a convergent functional integral, Horowitz 
\cite{ho:p} proposed an action in the form
\be
S=\int d^4 X\sqrt{g}[AC_{ijkl}^2+B(R-4\lambda)^2]
\ee
where $C_{ijkl}$ is the Weyl tensor, $R$ is the Ricci scalar, $\lambda$ 
is the cosmological constant and $A, B$ are the coupling constants. The 
action (1) reduces to the Einstein-Hilbert action at the weak energy 
limit \cite{ho:p}. To obtain a workable and simplified form of the field 
equations, one may consider a spatially homogeneous and isotropic 
minisuperspace background, for which the Weyl tensor trivially vanishes. 
In a previous paper \cite{a:p}, we considered the action (1) retaining 
only the curvature squared term. The field equations for such an action 
can be obtained by the standard variational principle. In the variational 
principle, the total derivative terms in the action are extracted and one 
gets a surface integral which is assumed to vanish at the boundary or the 
action is chosen in such a way that those surface integral terms have no 
contribution. However, for canonical quantization this principle is not 
of much help and one has to express the action in the canonical form, 
which is achieved only through the introduction of the auxiliary 
variable. Auxiliary variable can be chosen in an adhoc manner, and 
different choice of such variable would lead to different description of 
quantum dynamics, keeping the classical field equations unchanged. In 
view of this Ostrogradski \cite{o:g} in one hand and Boulware etal 
\cite{b:a} on the other, have made definite prescriptions to choose such 
variables. Ostrogradski's prescription \cite{o:g} was followed by Schmidt 
\cite{sc:p}. We shall consider Boulware etal's \cite{b:a} prescription in 
which it was proposed that the auxiliary variable should be chosen by 
taking the first derivative of the action with respect to the highest 
derivative of the field variable present in the action.
\par
Hawking and Luttrell \cite{h:n} utilized Boulware etal's \cite{b:a} 
technique to identify the new variable and showed that the 
Einstein-Hilbert action along with a curvature squared term reduces to 
the Einstein-Hilbert action coupled to a massive scalar field, assuming 
the conformal factor. Horowitz \cite{ho:p} on the other hand, showed 
that the canonical quantization of the curvature squared action yields 
an equation which is similar to the Schr$\ddot{o}$dinger equation. 
Pollock \cite{p:n} also used the same technique to the induced theory 
of gravity and obtained the same type of result  as that obtained by 
Horowitz \cite{ho:p}, in the sense that the corresponding 
Wheeler-deWitt equation looks similar to the Schr$\ddot{o}$dinger 
equation.
\par
The striking feature of Boulware etal's \cite{b:a} prescription is 
that it can even be applied in situations where introduction of the 
auxiliary variable is not at all required, eg. in the induced theory 
of gravity, vacuum Einstein-Hilbert action etc. It is not difficult to 
see that the classical field equations remain unchanged with or 
without the introduction of the auxiliary variables. Now the question 
is whether quantum dynamics remain unaffected? Putting other way 
round, the question turns out to be whether the solution of the 
Schr$\ddot{o}$dinger like equation obtained by Pollock \cite{p:n} in 
the induced theory of gravity by introducing auxiliary variable will 
satisfy the Wheeler-deWitt equation corresponding to the same action 
that can be obtained without introducing such variable? The answer is 
simply no. We have shown \cite{a:p} that the introduction of an 
auxiliary variable does not change the classical field equations in 
situation mentioned above, however, the quantum dynamics are 
different. The vacuum Einstein-Hilbert action in the background of 
Robertson-Walker metric has been chosen for this purpose. We have 
shown in such a toy model \cite{a:p}, `toy', since it is not 
required to follow Boulware etal's \cite{b:a} prescription in 
the vacuum Einstein-Hilbert action, that the introduction of an 
auxiliary variable leads to wrong Wheeler-deWitt equation. So 
what went wrong with Boulware etal's \cite{b:a} prescription? 
This is what we have concluded in \cite{a:p}, that before taking 
up the said prescription, one has to eliminate all possible 
total derivative terms from the action. Once this is done, it 
would not at all be possible to introduce auxiliary variables 
in situations mentioned above, where such introduction is 
unnecessary, and hence only a unique and correct quantum 
dynamics would emerge. In this sense, the auxiliary variable identified 
by Hawking and Luttrell \cite{h:n} and Schr$\ddot{o}$dinger like equation 
obtained by Horowitz \cite{ho:p} and Pollock \cite{p:n} 
are wrong. We \cite{a:p} have obtained the correct quantum dynamics 
for curvature squared action and showed that the 
continuity equation identifies the nature of space and 
time like variables in the minisuperspace, establishing, 
perhaps for the first time the quantum mechanical idea 
of the probability and the current densities in quantum 
cosmology. It has also been shown that the correct 
description of the curvature squared action leads to an 
effective potential, whose extremum yields vacuum 
Einstein's equation, which is a desirable feature of 
higher derivative gravity theory, in the weak energy 
limit. The emergence of the Schr$\ddot{o}$dinger like 
equation, the quantum mechanical concept of the 
probability density and the effective potential lead 
us to conclude that the quantum description of 
cosmology is incomplete without atleast a curvature 
squared term in the action. This is the reason why in 
this paper we have taken up the task of generalizing 
our previous work \cite{a:p} by modifying the 
Einstein-Hilbert action with a curvature squared term 
along with a matter field.
\par
As already mentioned, in order to obtain the correct 
and unique quantum cosmological description of a 
system whose action contains curvature squared term, 
one has to first remove all possible total derivative 
terms from the action and then follow Boulware etal's 
\cite{b:a} prescription. The whole formulation 
requires four steps, viz. 1) remove all possible 
total derivative terms from the action, 2) introduce 
auxiliary variable, which is the first derivative of 
the action with respect to the highest derivative of 
the field variable present in the action, 3) cast the 
action in the canonical form and obtain classical 
field equations, 4) introduce basic variables to 
canonically quantize the Hamiltonian constraint 
equation. At this stage we would like to mention  
that the Hamiltonian in our formalism before introducing basic variables 
\cite{a:p} turned out to be the same as 
that obtained by Schmidt \cite{sc:p} following 
Ostrogradski's \cite{o:g} prescription. This is due to the fact 
that, in that prescription \cite{o:g} total derivative terms do not 
appear. However, the importance of removal of the total derivative terms 
has not been stressed, the method of introducing auxiliary variables and 
process of casting the action in the canonical form is not so 
vivid and elegant as 
in Boulware etal's \cite{b:a} formalism and quantization procedure by 
introducing 
basic variables has not been mentioned in that approach \cite{o:g}. 
Therefore we have selected the other method.                          
\par
This paper has been organized in the following manner. In section 2. we 
present the field equation for an action containing curvature and 
curvature squared term along with a minimally coupled matter field, 
following the proposal mentioned above. Further we cast the action in 
canonical form with respect to the basic variables, which shows the 
appearence of a Hamiltonian $-P_{\alpha}$ of a system described by the 
basic variables and their conjugate momenta, though the actual 
Hamiltonian $H$ ie. the super Hamiltonian constraint to vanish. The form 
of the above action also identifies the time variable, which is the 
intrinsic time, whose canonical momentum is $P_{\alpha}$, or the 
Hamiltonian $-P_{\alpha}$ can be thought of as the generator of the 
translation along the above time variable. The action written in this 
form indicates the true dynamical degree of freedom, disentangled from 
the kinetic variable ie. the time variable. We claim that the Hamiltonian 
$(-P_{\alpha})$ is the correct Hamiltonian in the sense that it 
contains the basic variables and their conjugate momenta and its 
canonical quantization yields the Wheeler-deWitt equation which 
mimics the Schr$\ddot{o}$dinger equation, along with an effective 
potential. 
\par
In section 3. we present the conventional continuity 
equation of quantum mechanics from the Schr$\ddot{o}$dinger like 
Wheeler-deWitt equation. The continuity equation in quantum 
cosmology was first presented by Sanyal and Modak \cite{a:p} only in 
a recent publication and it is interesting to note that, as the 
probability and current densities are finite, so also the 
wavefunction and its derivatives are finite throughout the 
evolution of the Universe if and only if there are no singularities 
in the domain of quantum cosmology.
\par
Section 4. has been devoted to extract physical implication of the 
effective potential $V_{e}$. There are singularities in the 
effective potential. It is the function of the scalar field and 
the expansion parameter, which is the Hubble parameter of the 
classical field equations and it is diverging both at the 
zero's and at the very large values of the expansion 
parameter. The effective potential is asymetric with respect 
to the expansion and collapse of the Universe and its 
extremum yields classical equations. 
\par 
Section 5. is devoted in making some concluding remarks.  
\section{\bf{Classical field equations and quantization}}
We consider the following action
\be
S=\int d^4 x\sqrt{-g}[-\frac{1}{16\pi 
G}(R-\frac{\beta}{6}R^2)+\frac{1}{2\pi^2}(\frac{1}{2}\phi,_{\mu}\phi^{,\mu}-V(\phi))].
\ee
Under the following form of the closed Robertson-Walker metric
\be
ds^2=e^{2\alpha(\eta)}[d\eta^2-d\chi^2-sin^2 \chi(d\theta^2+sin^2 \theta 
d\phi^2)] 
\ee
the action (2) can be expressed as
\be
S=\int 
d\eta[\frac{3\pi}{4G}\{(1+\dot{\alpha}^2+\ddot{\alpha})e^{2\alpha}-\beta(1+\dot{\alpha}^2+\ddot{\alpha})^2\}+\frac{1}{2}\dot{\phi}^2 e^{2\alpha}-V(\phi)e^{4\alpha}],
\ee
where, $\dot{\alpha} = \frac{d\alpha}{d\eta}$. According to 
our proposal, we remove all removable total derivative terms 
from the action, so that only lowest order terms appear in 
the action (4). Thus,
\be
S=\int 
d\eta[M\{(1-\dot{\alpha}^2)e^{2\alpha}-\beta(1+\dot{\alpha}^2)^2-\beta\ddot{\alpha}^2\}+\frac{1}{2}{\dot{\phi}}^2 e^{2\alpha}-V(\phi)e^{4\alpha}]+S_{m},
\ee
where, $M = \frac{3\pi}{4G}$ and $s_{m} 
=M[\dot{\alpha}e^{2\alpha} - 2\beta(\dot{\alpha}+\frac{\dot{\alpha}^3}{3})]$.
\par
At this stage, we define the auxiliary variable, which is
\be
M\beta Q=-\frac{\partial{S}}{\partial{\ddot{\alpha}}}=2M\beta 
\ddot{\alpha},      ie. Q=2\ddot{\alpha}.               
\ee
Introducing the auxiliary variable $Q$ in action (5) and expressing the 
action in the canonical form, after removing the remaining total 
derivative terms, we obtain,
\be
S=\int 
d\eta[M\{\beta\dot{Q}\dot{\alpha}+(1-\dot{\alpha}^2)e^{2\alpha}-\beta(1+\dot{\alpha}^2)^2+\frac{\beta}{4}Q^2\}+\frac{1}{2}\dot{\phi}^2 e^{2\alpha}-V(\phi)e^{4\alpha}]+S_{m_{1}},
\ee
where, $S_{m_{1}}$ is the surface integral, given by $S_{m_{1}} = S_{m} 
- \beta Q\dot{\alpha}$. Assuming $S_{m_{1}}$ vanishes at the boundary, 
the classical field equations are found as, 
\be
\beta(\ddot{Q}-12\ddot{\alpha}\dot{\alpha}^2-4\ddot{\alpha})-2 e^{2\alpha} 
(1+\dot{\alpha}^2+\ddot{\alpha})-\frac{e^{2\alpha}}{M}(\dot{\phi}^2-4V(\phi)e^{2\alpha})=0,
\ee
\be
Q=2\ddot{\alpha},
\ee
\be
\ddot{\phi}+2\dot{\alpha}\dot{\phi}+e^{2\alpha} \frac{dV}{d\phi}=0,
\ee
\be
\beta(3\dot{\alpha}^4+2\dot{\alpha}^2-\dot{\alpha}\dot{Q}+\frac{Q^2}{4}-1)+(\dot{\alpha}^2+1)e^{2\alpha}=\frac{e^{2\alpha}}{M}(\frac{1}{2}\dot{\phi}^2+V(\phi)e^{2\alpha}),
\ee
the last one being the Hamiltonian constraint equation. The definition of 
$Q$ given in the equation (6) is thus recovered in equation (9). While 
equation (10) is the well known continuity equation of the scalar field $  
\phi$, equations (8) and (11) are the correct classical field equations 
as it is evident by replacing $Q$ by $2\ddot{\alpha}$. Now the Hamiltonian 
constraint equation in the phase space variables can be expressed as,
\be
H=\frac{P_{\alpha}P_{Q}}{M\beta} +\frac{P_{Q}^4}{M^3 
\beta^3}+\frac{P_{Q}^2}{M\beta^2}(2\beta+e^{2\alpha})+\frac{P_{\phi}^2 
e^{-2\alpha}}{2}+M\beta-Me^{2\alpha}-\frac{M}{4}\beta 
Q^2+V(\phi)e^{4\alpha}=0, 
\ee
where, $P_{\alpha}, P_{Q}$ and $P_{\phi}$ are the canonical momenta 
corresponding to $\alpha, Q$ and $\phi$ respectively, and $\dot{\alpha}, 
\dot{Q}$ and $\dot{\phi}$ are given by
\be
\dot{\alpha}=\frac{P_{Q}}{M\beta}, \dot{\phi}=e^{-2\alpha} P_{\phi}, 
\dot{Q}=\frac{P_{\alpha}}{M\beta}+\frac{2(2\beta+e^{2\alpha})}{M\beta^2}P_{Q}+\frac{4P_{Q}^3}{M^3 \beta^3}.
\ee
The Wheeler-deWitt equation is obtained through the quantization of the 
Hamiltonian constraint equation (12). Now as already mentioned, canonical 
quantization should be performed with the basic variables, viz. $\alpha$ 
and $\dot{\alpha}$. Let us choose $\dot{\alpha} = x$, and hence replace 
$P_{Q}$ by $M\beta x$, in view of the first equation given in (13). 
Further, from equation (6), we get 
\be
M\beta 
Q=-\frac{\partial{S}}{\partial{\ddot{\alpha}}}=-\frac{\partial{L}}{\partial{\ddot{\alpha}}}=-\frac{\partial{L}}{\partial{\dot{x}}}=-P_{x},
\ee
which means that $Q$ should be replaced by $-\frac{P_{x}}{M\beta}$. From 
equation (14) it is evident that $P_{x}$ is the canonical momentum 
corresponding to $x$. Hence equation (12) turns out to be
\be
H=xP_{\alpha}-\frac{P_{x}^2}{4M\beta}+\frac{1}{2}P_{\phi}^2 
e^{-2\alpha}+M\beta(1+x^2)^2+M(x^2-1)e^{2\alpha}+V(\phi)e^{4\alpha}.
\ee
As the Hamiltonian $H$ is constraint to vanish, we get 
\be
-P_{\alpha}=-\frac{P_{x}^2}{4M\beta x}+\frac{1}{2x}e^{-2\alpha} 
P_{\phi}^2+\frac{M\beta}{x}(1+x^2)^2+\frac{M}{x}(x^2-1)e^{2\alpha}+\frac{V(\phi)e^{4\alpha}}{x}.
\ee
We have already mentioned that $P_{\alpha},P_{\phi}$ and $P_{x}$ are the 
canonical momenta corresponding to $\alpha, \phi$ and $x$, respectively, 
however to understand the physical meaning of $P_{\alpha}$, we write the 
action (7) in terms of the canonical coordinates, apart from a surface 
term, remembering $H = 0$, as
\be
S=\int L d\eta=\int 
[P_{\alpha}\dot{\alpha}+P_{Q}\dot{Q}+P_{\phi}\dot{\phi}]d\eta 
\ee 
Further, from equation (14) we find
\be
\dot{Q}=-\frac{1}{M\beta}\frac{dP_{x}}{d\eta}=-\frac{\dot{\alpha}}{M\beta}\frac{dP_{x}}{d\alpha}.
\ee
Now using the expressions for $P_{Q}$ and $\dot{Q}$ given by equations (13) 
and 
(18) respectively, the relations $d\alpha = \dot{\alpha}d\eta, \dot{\phi} 
= \frac{d\phi}{d\alpha}\dot{\alpha}$ in action (17) and finally 
rearranging terms, we obtain
\be
S=\int 
[P_{x}\frac{dx}{d\alpha}+P_{\phi}\frac{d\phi}{d\alpha}+P_{\alpha}]d\alpha-\int 
\frac{d}{d\alpha}(xP_{\alpha})d\alpha. 
\ee
Apart from the surface term the action (19) looks similar to the 
conventional form of the action of a mechanical system described by the 
configuration space variables $x, \phi$ and their canonical conjugate 
momenta $P_{x}, P_{\phi}$, with a Hamiltonian $(-P_{\alpha})$, which is 
also the canonical momentum of $\alpha$ parametrized by the variable 
$\alpha$, though the true Hamiltonian $H$ of the gravitational system 
constraint to vanish. We know that the Hamiltonian of a system is the 
generator of the time translation. Since, $-P_{\alpha}$ in our system 
plays the role of the Hamiltonian, hence it acts as the generator of 
translation along $\alpha$, and the variable $\alpha$ acts as the time 
variable. This feature is also consistent with the intrinsic concept of 
general theory of relativity, as time has no independent existence from 
geometry in describing gravitation, rather it is inbuilt in the theory, 
unlike situations encountered in the conventional classical and quantum 
mechanics, where time is an external parameter. The functional form of 
the Hamiltonian $(-P_{\alpha})$ is given by (16) and it is a function of 
$x, \phi, P_{x}, P_{\phi}$ and the intrinsic time parameter $\alpha$. The 
canonical quanization of (16) yields
\be
i\hbar\frac{\partial{\psi}}{\partial{\alpha}}=-\hat{P}_{\alpha}
\psi=\frac{\hbar^2}{4M\beta x}(\frac{\partial^2 
{\psi}}{\partial{x}^2}+\frac{n}{x}\frac{\partial{\psi}}{\partial{x}})-\frac{\hbar^2}{2x}\frac{\partial^2{\psi}}{\partial{\phi}^2}e^{-2\alpha}+V_{e}\psi,
\ee
where $n$ is the operator ordering index and
\be
V_{e}=V_{e}(x,\phi,\alpha)=\frac{M\beta}{x}(x^2+1)^2+\frac{M}{x}(x^2-1)e^{2\alpha}+\frac{V(\phi)}{x}e^{4\alpha}. 
\ee
Equation (20) is the correct Wheeler-deWitt equation corresponding to an 
action, containinng curvature and curvature squared terms along with a 
minimally coupled scalar field. It looks similar to the 
Schr$\ddot{o}$dinger equation, though it differs from that obtained by 
Pollock \cite{p:n}.

\section{\bf{Probability and current density}}

One of the most important feature observed in quantum mechanics is that, 
the state of a system is described by a wavevector belonging to an 
abstract Hilbert space and the norm of the wavevector must be positive 
definite or zero. This idea emerged from the interpretation of the 
probability 
density to describe a given state from the continuity equation which is 
obtained by using the Schr$\ddot{o}$dinger equation. Probability 
interpretation follows from the simple mathematical appearence of the 
Schr$\ddot{o}$dinger equation. No such interpretation of the probability 
density in general exists in quantum cosmology, when the action contains 
terms linear in the Ricci scalar coupled with some matter field. This is 
due to the fact that there is no time a priori in the Wheeler-deWitt 
equation in a gravitational theory described by the action (2) with 
$\beta = 0$. It is to be noted that the continuity equation along with 
the conventional notion of the probability density can only be introduced 
with the proper choice of the canonical variables in such a way that the 
Hamiltonian constraint is quadratic in the canonical momenta along with a 
term linear in momentum whose canonical coordinate acts as an intrinsic 
time variable. This type of canonical quantization is possible only when 
the action contains atleast a quadratic curvature term.
\par
Equation (20) can be writtn as    
\be
i\hbar\frac{\partial{\psi}}{\partial{\alpha}}=\hat{H}_{0}\psi,
\ee
where
\be
\hat{H}_{0}=\frac{\hbar^2}{4M\beta 
x}(\frac{\partial^2}{\partial{x}^2}+\frac{n}{x}\frac{\partial}{\partial{x}})-\frac{\hbar^2}{2x}e^{-2\alpha}\frac{\partial^2}{\partial{\phi}^2}+\hat{V}_{e}(x,\phi,\alpha).
\ee
It is to be noted that $\hat{H_{0}}$ operator is hermitian and as a 
consequence we obtain the continuity equation, viz.
\be
\frac{\partial{\rho}}{\partial{\alpha}}+\bf{\nabla}\cdot{\bf{J}}=0,
\ee
where $\rho$ and $\bf{J}$ are the probability and the current densities 
respectively for the choice of the operator ordering index $n = -1$. 
$\rho$ and $\bf{J}$ are given by $\rho = \psi^*\psi$ and ${\bf{J}} = (J_{x}, 
J_{\phi}, 0)$, where
\be
J_{x}=\frac{i\hbar}{4M\beta x}(\psi^*\psi_{x}-\psi\psi_{x}^*),
\ee
\be
J_{\phi}=-\frac{i\hbar 
e^{-2\alpha}}{2x}(\psi^*\psi_{\phi}-\psi\psi_{\phi}^*), 
\ee
and
\be 
{\bf{\nabla}}=(\frac{\partial}{\partial{x}},\frac{\partial}{\partial{\phi}},0).
\ee
One can also find the continuity equation for other values of the 
operator ordering index but with respect to a new variable $y$ which is 
functionally related to $x$ only. It should be mentioned here that the 
continuity equation in quantum cosmology was presented by Sanyal and Modak 
\cite{a:p} for the first time in a recent publication. In the 
present work we have 
obtained the continuity equation once again. Since the above probability 
and the current densities are dynamical quantities, therefore the 
wavefunction and its derivatives should remain finite at all epoch of the 
evolution of the Universe, if and only if there are no singularities in 
the domain of quantum cosmology.
\par
Further in analogy with the quantum mechanics it is noted that the 
variable $\alpha$ in equation (20) can be identified as the time 
parameter in quantum cosmology. The variables $x (=\dot{\alpha})$ and 
$\phi$ act as spatial coordinate variables in the context of quantum 
cosmology. The coordinate variable $x(=\dot{\alpha})$, which is just 
the expansion parameter and the Hamiltonian operator $\hat{H}_{0}$ 
diverges at the bounce of the Univese, ie. at $\dot{\alpha} = 0$.      

\section{\bf{Effective potential and its extremization}}
The effective potential given by equation (21) is found to be assymetric 
with respect to the expansion parameter $(x)$, ie. $V_{e}(-x) = -V_{e}(+x)$ 
and it is also unstable both at the short range and the long range values 
of $x$.The significance of the effective potential becomes clear when 
it is extremized, yielding interesting results. In the weak energy 
limit, ie. when the deBroglie wavelength is much larger in comparison 
with the average separation (ie. far below the Planck scale) the 
contribution of the kinetic energy is sufficiently small with respect 
to the potential energy in the Hamiltonian . At this regime the 
Hamiltonian is almost dominated by the potential energy. Now the 
extremum of the effective potential $V_{e}$ given by equation (21), 
with respect to $x$ yields, keeping $\phi$ and the time parameter 
$\alpha$ fixed,
\be
\beta(3x^4+2x^2-1)+(x^2+1)e^{2\alpha}-\frac{e^{4\alpha}}{M}V(\phi)=0.
\ee
If $V(\phi)$ has got no extremum then in the limit $\beta \rightarrow 
0$, ie. in the absence of the curvature squared term, equation (28) 
takes the following form
\be
x^2+1=\dot{\alpha}^2+1=\frac{e^{2\alpha}}{M}V(\phi),
\ee     
which is exactly the classical constraint equation in the Einstein's 
gravity, along with a scalar field, apart from a kinetic energy part 
corresponding to the scalar field, that we have already assumed to be 
small. Further, for the action dominated by the curvature squared term, 
ie. at $\beta \rightarrow \infty$, we have
\be
x^2+1=0 ,
\ee
or,
\be
x^2-\frac{1}{3}=0.
\ee
These pair of equations have already been obtained in \cite{a:p}. Here, 
we have just been able to recover them in the limit $\beta \rightarrow 
\infty$, ie. at the epoch of the evolution of the Universe dominated by 
the curvature squared term. These equations imply that the extremum of 
the effective potential $V_{e}$ is at a coordinate position $x = 
\dot{\alpha}$, satisfying either equation (30), which is the vacuum 
Einstein's equation that admits Euclidean wormhole solution, or equation 
(31), which is the epoch at which the Universe admits a solution, given 
by $a = \frac{(t - t_{0})}{\sqrt{3}}$, `a' being the scale factor in 
proper time `t'. For this solution the horizon radious $r_{H}$ is 
proportionaql to $ln(t)$, and so as $t \rightarrow 0, r_{H} \rightarrow 
\infty$, and thus the horizon problem is solved. The effective 
potential $V_{e}$ diverges in both the directions of $x \rightarrow 0$ 
and $x \rightarrow \infty$, even for $\beta \rightarrow \infty$. 
Further, it can be shown that $\frac{\partial^2{V_{e}}}{\partial{x^2}} > 
0$, for condition (31), ie. $V_{e}$ has got a minimum. These 
altogether imply that the Universe, while sitting at the minimum of 
the effective potential, will expand at the rate $a$ proportional to 
$t$, solving the horizon problem. 
\par
Now, had there been an extremum of the scalar field potential 
$V(\phi)$, then extremizing the effective potential $V_{e}$, given 
in equation (21), with respect to $\phi$, keeping $x,\alpha$ 
fixed, we get
\be
\frac{e^{4\alpha}}{Mx}\frac{dV}{d\phi} = 0.
\ee
The condition that a function $(V_{e})$ of two variables 
$x$, and  $\phi$ (not considering the time parameter $\alpha$) has got 
a minimum, is
\be
\frac{\partial^2{V_{e}}}{\partial{x^2}}\frac{\partial^2{V_{e}}}{\partial{\phi^2}}-\frac{\partial^2{V_{e}}}{\partial{x}\partial{\phi}} > 0.
\ee
The left hand side in the present context turns out to be
\be
\frac{2e^{4\alpha}}{Mx_{0}^2}[e^{2\alpha}+2\beta(3x_{0}^2+1)]\frac{d^2 
 V(\phi)}{d\phi^2}|_{at \phi=\phi_{0}}
\ee
which is positive definite for $\frac{d^2 V(\phi)}{d\phi^2}|_{at 
\phi=\phi_{0}} > 0$, ie. for a minimum of the scalar field 
potential. Therefore the effective potential has got an extremum 
located at the configuration space variables $\phi = \phi_{0}$ and 
$x = x_{0}$, determined by equation (28).
\par
It can further be shown that $\frac{\partial^2 
V_{e}}{\partial{x^2}}|_{\phi,\alpha=constant} > 0$, which means that 
the effective potential is truely a minimum at the location of the 
configuration space variables $\phi = \phi_{0}, x = x_{0}$, 
determined by equation (28), where $V(\phi) = V(\phi_{0})$. 
Further, if we choose the minimum of the scalar field potential 
$V(\phi_{0}) = 0$, then equation (28) reduces to either 
\be
x^2+1=0
\ee
or,
\be
\beta(3x^2-1)+e^{2\alpha}=0.
\ee
these pair of equations again imply that the Universe is sitting at 
the minimum of the effective potential $V_{e}$, at the locations of 
the configuration space variables $\phi_{0}$ and $x = x_{0}$, 
satisfying either equation (35), which is the vacuum Einstein's 
equation admitting Euclidean wormhole solution, or equation (36), 
that admits an oscillatory solution, in the form
\be
a^2=\beta sin^2(\frac{t-t_{0}}{\sqrt{3}\beta}),
\ee
where, as already mentioned, `a' is the scale factor in proper 
time `t'.            

\section{\bf{Concluding remarks}}
In summary, we have observed that, different choice of the auxiliary 
variable, leads to different inequivalent quantum dynamics, keeping the 
classical field equations unchanged. Hence, we have suggested a principle 
of choosing the correct auxiliary variable and the true degree of 
freedom, which would lead to the correct and unique quantum description. 
For this we have modified Boulware etal's principle \cite{b:a} by adding 
only one statement at the begining that, before choosing such 
variables one has to remove all removable total derivative terms 
from 
the action. That this would lead to the correct quantum description, has 
been proved in a toy model, in a recent publication by Sanyal and Modak 
\cite{a:p}. In addition, it has been shown that the quantum dynamics 
obtained by this principle leads to certain desirable features viz. 
quantum mechanical probability and current density interpretation from 
the continuity equation of quantum cosmology and an effective potential 
whose extremum yields classical field equations. It seems that a quantum 
mechanical probability interpretation is possible, only if the 
Einstein-Hilbert action in the Robertson-Walker minisuperspace model is 
modified by atleast curvature squared term, 
which appears in the one loop correction of perturbative quantum field 
theory in curved space time, and is the most dominant term in the 
quantum domain. Thus we may conclude that, quantum mechanical 
probability interpretation of quantum cosmology is a generic feature of 
curvature squared gravity, at least in the Robertson-Walker 
minisuperspace model.

\end{document}